\documentstyle[eqsecnum,aps,amstex,epsfig,float]{revtex}

\begin{document}                
\draft
\title{An efficient filter for detecting gravitational wave bursts in 
interferometric detectors}

\author{Thierry Pradier, Nicolas Arnaud, Marie-Anne Bizouard, Fabien Cavalier, Michel Davier and Patrice Hello}

\address{ Laboratoire de l'Acc\'el\'erateur Lin\'eaire, B.P. 34,B\^atiment 200,
Campus d'Orsay,
91898 Orsay Cedex (France)\protect\\}

\maketitle
\begin{abstract}
Typical sources of gravitational wave bursts are supernovae, for which no accurate models exist. 
This calls for search methods with high efficiency and robustness to be used in the data analysis
of foreseen interferometric detectors. A set of such filters is designed to detect gravitational 
wave burst signals. 
We first present filters based on the linear fit of whitened data to 
short straight lines in a given time window and combine them in a non linear filter
named ALF. We study the performances and efficiencies of these
filters, with the help of a catalogue of simulated supernova
signals. The ALF filter is the most performant and most efficient
of all filters. Its performance reaches about 80\% 
of the Optimal Filter performance designed for the same
signals. Such a filter could be implemented as an online
trigger (dedicated to detect bursts of unknown waveform)
in interferometric detectors of gravitational waves.

\end{abstract}




\section{ Introduction}
Long baseline interferometric detectors of gravitational waves (GW) 
\cite{ligo,virgo,geo,tama}
will be operational in the next years. The preparation for data analysis with 
these new instruments has begun since a long time for the compact binary 
inspirals, the most promising source of GW to date, and for periodic sources 
as well (see eg \cite{schutz} for a review). 
On the other hand, it is important to develop analysis methods to search for GW
bursts for which no accurate models exist. 
Typical sources of GW bursts are supernovae (historically the first cosmic 
sources of gravitational radiation ever considered). Simulations 
of collapses of isolated massive stars to neutron stars (type II supernovae) 
\cite{monchmeyer,bona1,zwerger,rampp} suggest small departures from spherical
symmetry. As a consequence, the power radiated away by GW 
during the few milliseconds of the collapse remains very low : the typical GW 
amplitude expected for such a source located at 10 Mpc
does not exceed $10^{-22}$-$10^{-23}$. 
This seems to give only hope for detecting supernovae 
events from inside our Galaxy, given the expected initial sensitivity of the
current projects. Collapses of more massive stars to black holes don't seem to 
provide much larger amplitudes of GW \cite{stark}. 
One important aspect is that these simulations 
are unable to predict accurate waveforms for the signals, as a small change 
in parameters can completely change the shape of the waveform (see \cite{zwerger} 
for example). This situation calls for search methods with high efficiency and
robustness against waveform variations.

Mergers of compact binaries \cite{ooh} can also be considered as burst sources 
with perhaps good chances of detection.
Some recent estimates for the amplitude of GW during the merging of two neutron 
stars give numbers as high as a few $10^{-21}$ for sources located at 10 Mpc \cite{ruf}. 
The merging of a neutron star and and black hole seems even more efficient 
with amplitudes near $10^{-20}$ for sources at 10 Mpc \cite{janka}. The
predicted amplitudes are just above the noise level of initial interferometric 
detectors, hence a likely detection.
Note also that these two 
kinds of merging compact binaries are likely to be also strong emitters of 
gamma-ray bursts and, so studies of coincidences between GW detectors and 
gamma-ray burst detectors on satellites can be crucial to validate the GW
detection \cite{FMR}. Here again, 
the details of the waveforms in the merging phase are poorly predicted.
Finally, concerning the merging of two black holes, the Binary Black Hole Grand 
Challenge Alliance \cite{gca} intends to compute numerically the waveforms 
emitted during black hole collision and coalescence.
Recent results suggest that GW amplitudes could also be of the order of
a few $10^{-21}$ for a total binary mass around 10 $M_\odot$ located
at 10 Mpc \cite{kgpp}.

Sources of GW bursts are then characterised by poor predictions of the 
emitted waveforms. At best, we only have ideas about bandwidths or typical 
frequencies of the signals. Matched filtering, as used for the 
detection of inspiralling binaries is clearly ruled out in this case and
robust methods for detecting this kind of sources are then required.

Some methods have been recently proposed and studied. The ``power 
filter'' technique has been introduced by Flanagan and Hugues
\cite{flana} in the context of binary black hole mergers, and developed
further by Anderson {\it et al.} \cite{powermonit,power2}. The idea here is to monitor the 
noise power along the time; it can be shown that this filter is optimal when 
only signal duration and bandwidth are known \cite{powermonit,power2}. A similar idea 
(``Norm Filter'') has been tested independently by Arnaud {\it et al.} \cite{arnaud}
to detect supernova GW signals.
Time-frequency methods \cite{bala} should be also pertinent for detecting 
unmodelled bursts; because of their computing costs, these methods are more 
suited to the off-line (re-)analysis of candidates selected by faster online 
algorithms. Of course, one can hardly distinguish between a real burst GW signal
and a transient burst caused by noise; thus,  methods devoted to detect non 
stationarity in the noise\cite{mohanty} are then able to detect ``true'' 
signals as well. Conversely, general filters are sensitive to
transient noises as well as to bursts signals. If selected by on-line triggers,
these spurious events can be  eliminated if they are coincident with signals
detected in auxiliary sensors sensitive to different kinds of environmental
noise (seismic activity, RF pickup, ...). Otherwise they can be validated
when searching for coincidences
between candidates from different GW detectors \cite{jolien}. Furthermore,
if an event is seen in coincidence by, say the three
interferometers of the LIGO-Virgo network, then it will be possible
to reconstruct the characteristics of the emitted GW signal \cite{gursel}. 
In particular, the 
reconstruction of the location of the source in the sky will permit
to search for  further
coincidences with other types of detectors (optical telescopes or
gamma-ray satellites for example), and thus enhance the confidence level
of the detection.

Our purpose is to develop and test filters for GW burst detection 
which are efficient, yet simple and 
fast enough to be used as on-line triggers\cite{arnaud,moriond,pra}. 
In this paper, we 
propose to study a family 
of filters based on slope detection algorithms, similar to existing 
contour detection algorithms used for image processing, applied here 
to the simpler one-dimensional case. The basic idea is to detect a 
non zero slope in the data stream delivered 
by interferometric detectors. In a first step the data are 
whitened by some suitable procedure \cite{cuo,cuo2}, so that we assume  that
the noise is Gaussian and white. If we fit a finite-length time series containing
only noise to a straight line, a null slope and a null offset are
obtained. A non vanishing slope could then
indicate the presence of some signal added to the noise. In the following, 
we will first study as filters for detecting GW bursts, the two parameters of 
a linear fit, namely the slope (slope detector) and the offset 
(offset detector). These two filters are strongly correlated;  it is however possible
to decorrelate them, and finally 
combine them in an unique filter using the complete information.
Next, we compute the performance of 
these filters, following a procedure already described in \cite{arnaud},
and compare them to filters previously tested 
\cite{arnaud} and to the optimal filter taken as reference. 
We finally study the efficiency of the filters (fraction of events detected 
for a given signal over many different noise realisations).

\section{Slope detection and related filters}

\subsection{The noise model}
Throughout the paper, we assume that the noise is Gaussian and white with zero 
mean. The standard deviation of the noise is then :
\begin{equation}
\sigma = \sqrt{ S_h f_0 \over 2 },
\label{sigma}
\end{equation}
where $f_0$ is the sampling frequency and $S_h$ is the one sided spectral 
density of the noise \cite{corr}.
For numerical examples, we take $f_0=20$ kHz (Virgo sampling rate)
and $\sqrt{S_h} \simeq 4\times 10^{-23}$ $/\sqrt{\mathrm{Hz}}$, which is about 
the minimum value of the foreseen noise spectral density of the Virgo 
interferometer \cite{virgosens}; this choice is correct since the minimum is lying right in the frequency 
range for expected burst sources of GW. The fact that we choose
a Gaussian noise is not essential, but simply convenient for the design 
of the filters. Deviation from gaussianity 
will produce for example an excess in the rate of false alarms and it will then
be possible 
to retune the algorithms according to the real noise statistics. 
In the frequency
range of interest, above a few 100 Hz,
 the Virgo noise sensitivity curve is rather flat, although 
not exactly white. The filtering methods presented here and in \cite{arnaud} 
require a whitening of the noise \cite{cuo,cuo2}, which is foreseen for the
output of the Virgo data. In the following, we normalise the 
noise by its standard deviation, so that we are dealing with a Gaussian noise 
with zero mean and unit standard deviation.
\subsection{Detecting a non zero slope in the data}

Let us divide the data set in sliding time windows with N samplings.
Fitting the data $h(t)$ to a straight line $at+b$, 
we obtain the slope $a$ and the offset $b$,
\begin{align}
a &= {<th>-<t><h> \over <t^2>-<t>^2}, \\
b &= <h>-a<t>,
\end{align}
where $<x>= {1\over N} \sum_{i=1}^N x_i$ is defined as the arithmetic mean of
the $x_i$ in the interval of length $N$. Here, $t_i=i/f_0$ is the $i^{th}$ 
sampled time and $h_i$ is the $i^{th}$ sampled value of the detector output. 
The fit transforms the input normalised Gaussian noise into 
Gaussian random variables, considered as linear filters, 
with zero mean and standard deviations $\sigma_a$ and $\sigma_b$ given by :

\begin{align}
\sigma_a^2 &=  {12 f_0^2 \over N(N^2-1)}, \label{sa} \\
\sigma_b^2 &=  {4N+2 \over N(N-1)}.
\end{align}

We can finally compute the signal to noise ratio (SNR) for each of the two 
filters, $X_a = a/\sigma_a$ and $X_b = b/\sigma_b$, noting
that the only free parameter for both filters 
is the length of the analysis window $N$. In practice, their implementation is 
very simple and fast thanks to trivial recursive relations between filter 
outputs for two successive windows.

\begin{figure}
\centerline{\epsfig{file=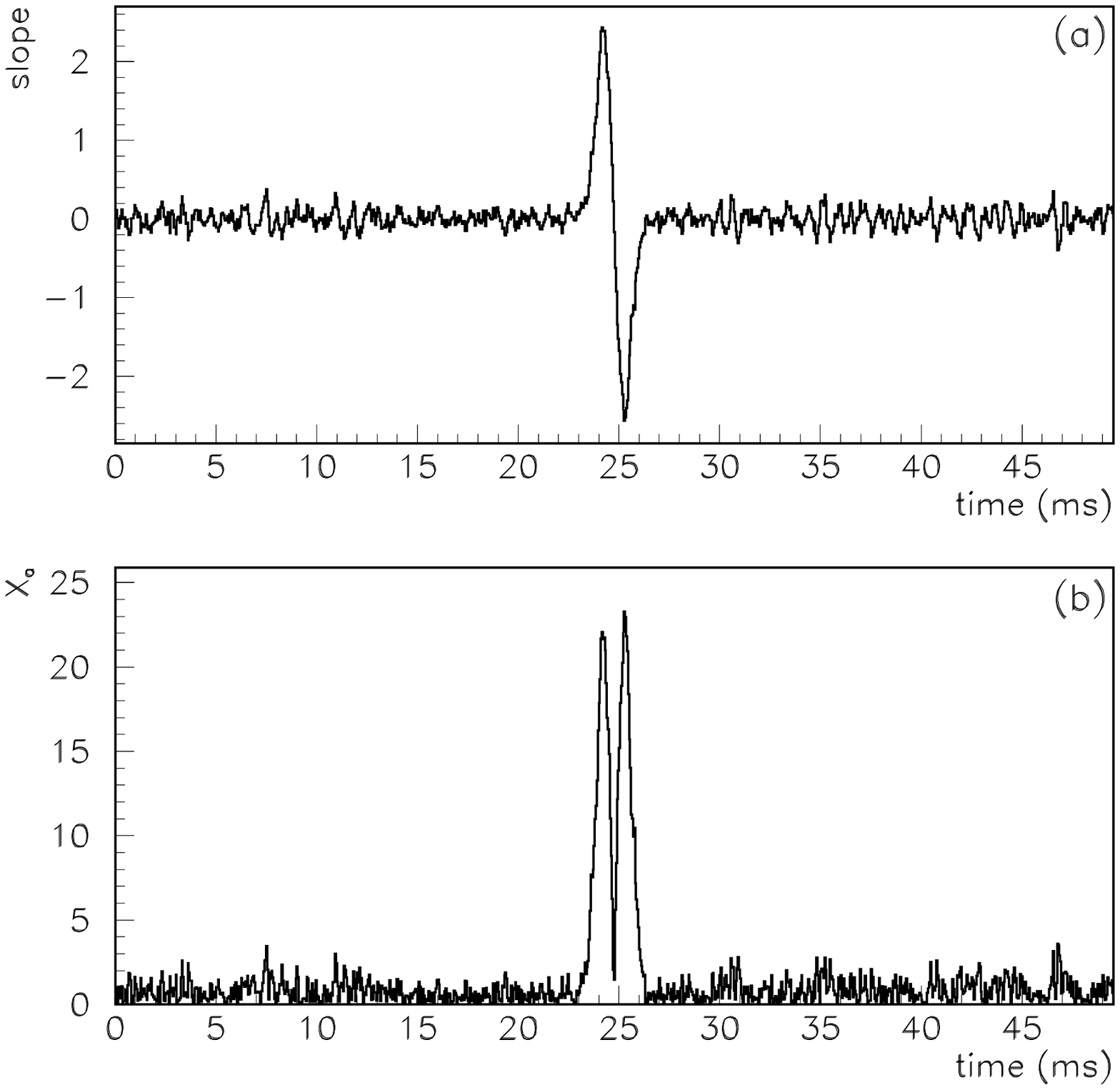,height=10cm}} 
\caption{ Slope $a$ (upper) and SNR $X_a$ (lower) for a Gaussian burst signal of 
width $\Delta=$0.5 ms. The analysis window size is $N=10$, {\it i.e} 0.5 ms. }
\label{slope1}
\end{figure}

\begin{figure}
\centerline{\epsfig{file=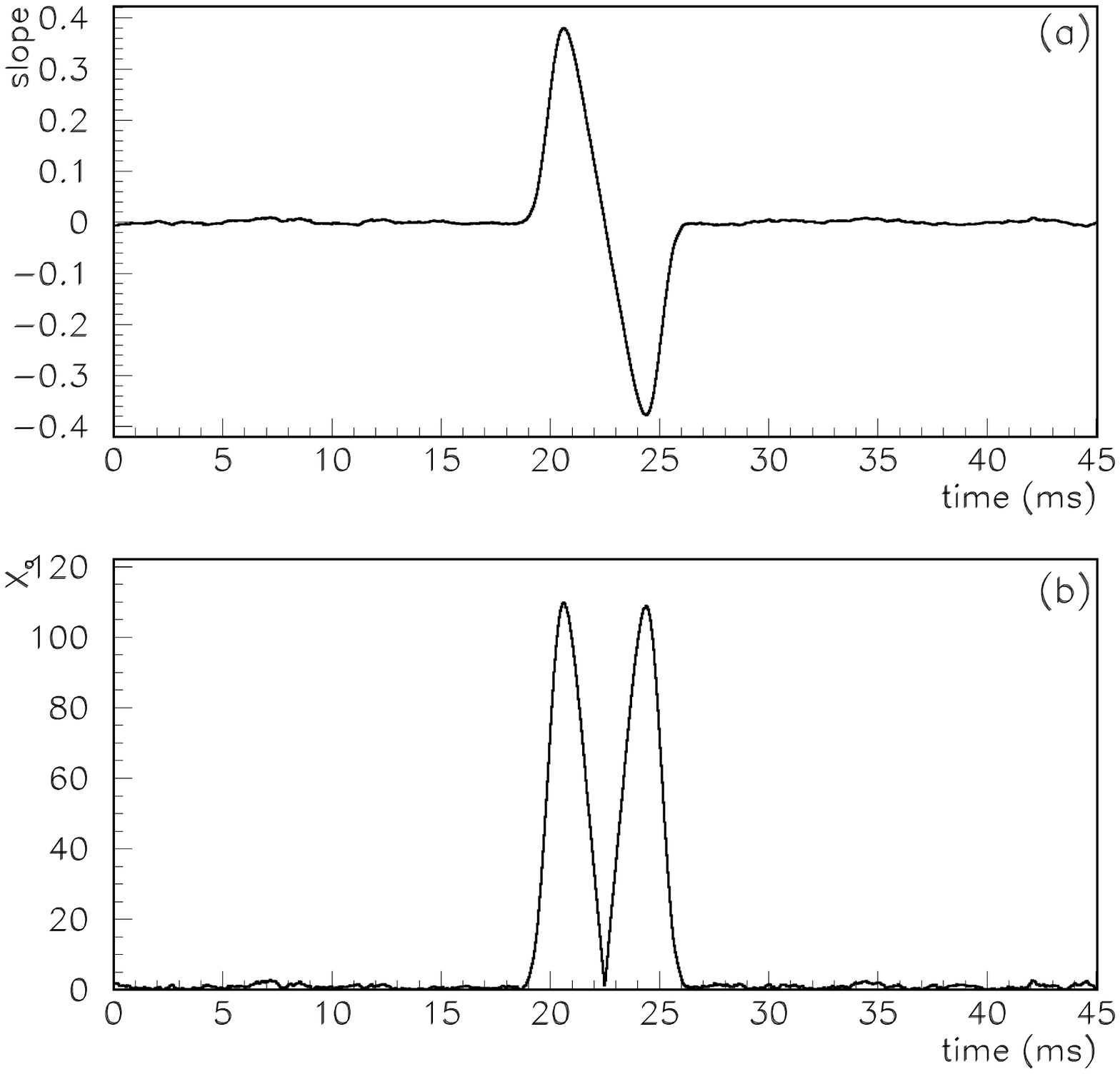,height=10cm}} 
\caption{ Slope $a$ (upper) and SNR $X_a$ (lower) for a Gaussian burst signal of 
width $\Delta=$0.5 ms. The analysis window size is $N=100$, {\it i.e} 5. ms.}
\label{slope2}
\end{figure}

\begin{figure}
\centerline{\epsfig{file=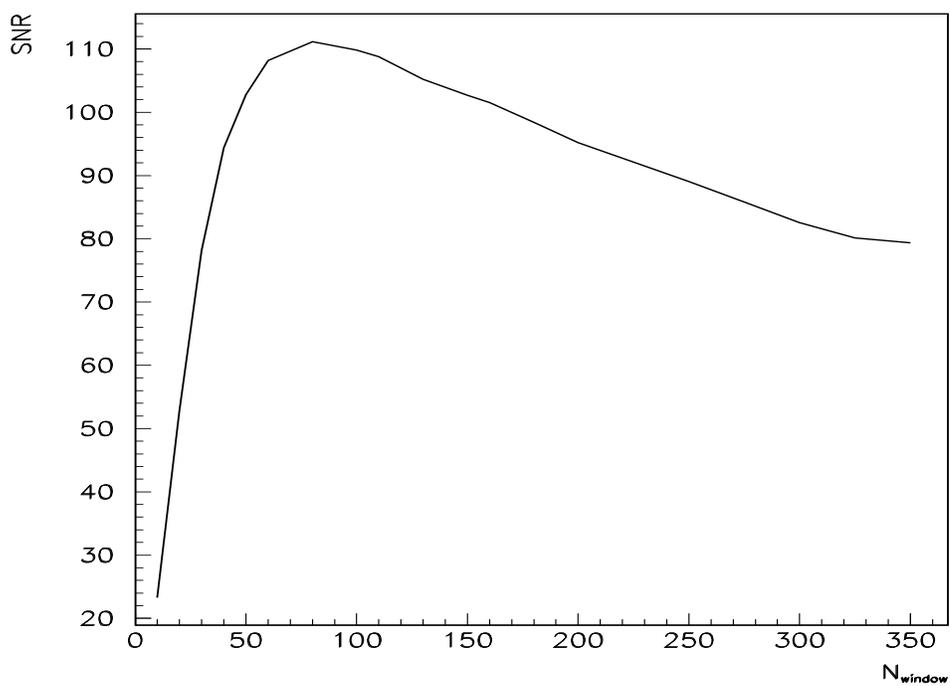,height=14cm,angle=270}} 
\caption{SNR $X_a$ for a Gaussian burst signal of 
width $\Delta=$0.5 ms as a function of the analysis window size $N$.
The maximum SNR is obtained for $N\simeq 7\Delta$.}
\label{slope3}
\end{figure}

It is interesting to notice that the maximum 
SNR $X_a$ or $X_b$ with respect to the window
size $N$ is in general not obtained when the slope $a$ or the offset $b$ 
is maximum. 
This point is illustrated in Fig. \ref{slope1} and 
\ref{slope2}, where we plot the slopes and SNR for
a Gaussian burst signal $\exp(-(t-t_0)^2/2\Delta^2)$ with $\Delta=0.5$ ms
and $t_0=25$ ms (signal maximum at the centre of the time scale). 
In Fig. \ref{slope1}, the analysis window size is $N=10$, while in Fig. \ref{slope2}, it
is $N=100$. The slope computed by the fit procedure is much larger 
in the first case ($N=10$) than in the second ($N=100$)(factor
about 5). Nevertheless, the SNR is higher (by a factor
about 5 in this example) for the $N=100$ window than for the $N=10$ window. This is
due to the fact that a larger window
size allows to average the effect of the noise; indeed, from Eq. (\ref{sa}),
we see that $\sigma_a$ scales as $1/N^3$. Thus, for a Gaussian signal of width
$\Delta$, the optimal window size is found to be about $7\Delta$, as seen
in Fig. \ref{slope3}.
The same is observed for the offset detector (to a lesser extent) and
for the derived  filters described below.

\subsection{Decorrelation of the slope and offset detectors}

In case of noise alone, the normalised offset and slope detectors $X_a$ and 
$X_b$ are two highly correlated random variables. 
They can be decorrelated by 
diagonalising the covariance matrix of $X_a$ and $X_b$ :
\begin{equation}
C = \begin{pmatrix} 	1 & \alpha \\ 
	    		\alpha & 1  	\end{pmatrix}
\end{equation}
where $\alpha$ = cov$(X_a,X_b)$. The eigenvalues of $C$ are then $1\pm\alpha$, 
corresponding to the eigenvectors $X_a\pm X_b$. Two new uncorrelated
random variables are introduced :
\begin{equation}
X_\pm = { X_a \pm X_b \over \sqrt{2(1\pm\alpha)} }
\end{equation}

$X_\pm$ are normalised in such a way that they are standard normal
variables, if $X_a$ and $X_b$ are standard normal variables.

The computation of the covariance $\alpha$ is easy, yielding :

\begin{equation}
\alpha =  -\sqrt{{ 3\over2} \left({N + 1\over 2N + 1 }\right)}.
\end{equation}

The two new statistics $X_\pm$ can be used as uncorrelated 
filters for detecting GW bursts, but
can also be easily combined into an unique filter.

\subsection{The combined filter : ALF}

The optimal variable retaining the full information contained in the
slope and offset filters is :
\begin{equation}
A=X_+^2+X_-^2 = {X_a^2+X_b^2-2\alpha X_aX_b \over 1-\alpha^2}. 
\end{equation}
In the absence of signal in the noise, $A$ is well approximated by a $\chi^2$ with 2 degrees
of freedom, as a sum of the square of two uncorrelated (albeit not independent) normal variables. 
The filter based on $A$ is called ALF (Alternative Linear fit Filter). Again, the only free parameter 
is the window size $N$. Note that ALF is not a linear filter, contrary to  the slope, offset  
and $X_\pm$ filters.

\subsection{Threshold for detection and false alarms}

\subsubsection{Redefinition of an Event}

A major problem arises when such filters are implemented, and tested with real (or simulated) data. To be optimally efficient and because of the short durations of the signals we are interested in, each filter is applied every time step ($\delta t = 5 \times 10^{-5} s$ for Virgo). As a consequence, a same false alarm is likely to appear for different window sizes, if different analysis windows are used in parallel, and for consecutive windows : this is the multi-triggering problem.

The redefinition of an {\it event} solves this problem. For each window size $N$ used in the implementation of the filters, we note $t_{start}$ and $t_{end}$ the time characteristics of each triggered event (time serie of data points for which SNR $\geq \tau$, where $\tau$ is the detection threshold). One has to take into account all the possibilities of overlapping between the different intervals. For instance, $[t_{start1}, t_{end1}]$ (for analysis window $N_1$) and $[t_{start2}, t_{end2}]$ (for analysis window $N_2$) will describe the same {\it event} if, e.g, $t_{start1} \leq t_{start2} \leq t_{end1}$ and $t_{end1} \leq t_{end2}$. Each selected event will be a cluster of points, characterised by a starting time and an ending time. 

\subsubsection{General discussion}

We set a detection threshold by choosing a false alarm rate $\kappa_0$. In all 
the following numerical examples, we consider $\kappa_0 = 10^{-6}$, 
corresponding to 72 false alarms per hour for
a 20 kHz sampling frequency. This choice results from a compromise
between the necessary data reduction after online processing (which should not
exceed a few percent) and the weakness of the GW signals we are 
looking for. For instance, it will be shown in the following that, with such a
false alarm rate, optimal filtering of a sample of supernovae signals gives, on average,
an upper limit for the distances of detection of the order of the radius 
of our galaxy, using realistic simulations for the emitted waves.
A large part of these false alarms will be 
in principle then discarded, when working afterwards in coincidence with 
other detectors, supposing that the different detectors noises are well
uncorrelated. Obviously, this rate should be adjusted in future coincidence
experiments by the maximum allowed rate for accidental coincidences.
One could have chosen a false alarm rate so that the mean detection distance obtained by Matched Filtering of realistic supernovae waveforms corresponds to, e.g, the diameter of the Milky Way ($R \simeq 30\mbox{ kpc}$) or the distance of the Magellanic Clouds ($d \simeq 55\mbox{ kpc}$). In both cases, however, the number of false alarms is too high to be manageable (hundreds or thousands of false alarms per hour). 
 
Anyway, the exact choice of the
false alarm rate, and the corresponding threshold, is not important 
in this study, since
we concentrate in the following on the {\it relative} performances of the 
filters (relative with respect to the optimal filter with the same
false alarm rate), provided that these performances do not crucially depend on the false alarm rate. Fig. 4 shows the evolution of the relative performance of the ALF filter averaged over a sample of
realistic supernovae signals (described below) as a function of the false alarm rate. This performance shows to be relatively constant for weak false alarm rates, and begins to increase for (non relevant) extremely high false alarm rates ($\kappa \simeq 10^{-3}$ corresponds to several thousands of false alarms per hour) ; this last feature is due to the redefinition of a event, which gives a larger threshold reduction for large false alarm rates.
 
The false alarm rate chosen in this paper corresponds to a threshold
of about $\tau \simeq 4.89$ for a normalised Gaussian variable (slope and 
offset detectors,
 $X_+$ and $X_-$) and to a threshold of about $27.63$ for a 
two-dimensional $\chi^2$ (ALF). This supposes of course an implementation
with an unique window size ; if several window sizes are to be used in parallel, 
then the threshold has to increased accordingly to keep the same overall
false alarm rate. The actual false alarm rate is then altered by the redefinition of 
events that is adopted here. As an example, for a single-windowed Slope
Detector, the same detection threshold would correspond to false alarm rate $\kappa_0 = 10^{-6}$ if
the definition of an event is taken into account, and to a false alarm rate roughly equal to $2 \times \kappa_0$ 
if not.

\subsubsection{Can we use the clustering information ?}

The real false alarm rate will correspond to the number of streams of data points $N_i$ in which 
$SNR \geq \tau$ ($\tau$ is the detection threshold). The information provided by clustering can be 
used in two different ways. 

First, for a given false alarm rate $\kappa$ (corresponding to a detection threshold $\tau$ for ALF), and for
 a given window size, one can determine the probability P(cluster size $\geq$ $n$) 
for a cluster of size larger than $n$  to occur. A cluster of size larger than
$n$ will then be found with a rate $\kappa \times P(n)$. An integer
$n_0$ can be found such that $\kappa \times P(n_0) = \kappa_0$. As a consequence, 
the detection threshold $\tau$ for ALF corresponding to $\kappa$ will be lowered.
 The definition of an event in this case will thus be a cluster of size $\geq$ $n_0$ for which SNR 
$\geq \tau$. The results obtained with such a definition are similar to the ones described in the next 
section.

The distribution of the number of consecutive triggers for a given threshold can be used in another way. This distribution gives a probability of occurrence $P(n)$ of a given number $n$ of consecutive triggers for the fixed threshold $\tau$. Then, each cluster can be labelled with the corresponding probability, and one can put {\it priority} in the treatment of those events. Furthermore, putting a threshold on the quantity $n$ can help to remove some of the false alarms. In this case, one can hope to discard a substantial part of the events, which with great probability are actually false alarms. Of course, the loss of signal this process causes has to be quantified. For high SNRs physical signals, using ALF, an average loss as high as $20 \%$ is observed for a $50 \%$ false alarms removal. The price to pay for such a removal is clearly too high.

\section{Detection of supernovae}
\label{section detection}

In order to benchmark the filters, we use a catalogue of simulated supernova GW
signals. Indeed, as we need ``robust'' filters with respect to the details
of the waveforms, it seems convenient to average the filters performances
over a variety of (physically sound) waveforms.

\subsection{The catalogue of signals}

As in \cite{pra} and \cite{virgonotes}, we use as a catalogue the 78 GW signals simulated by Zwerger and M\"uller 
\cite{zwerger} which are available in \cite{webSN}. These waveforms
result from the collapse of massive stars into neutron stars within the
assumption of axial symmetry. Each signal corresponds to a particular set
of parameters, mainly the initial distribution of angular momentum inside
the progenitor star and the rotational energy in the core. 
All the signals are computed for a source located at
10 Mpc; we can then re-scale the amplitudes of waveforms in order to locate the source
at any distance. In the following, we assume that the incoming waveforms
are optimally polarised, along the interferometer arms; this assumption
has no consequence on the relative performances of the filters. The detection 
distances displayed below have then to be considered as upper limits.
Zwerger and M\"uller distinguish three different types of waveforms.
Type I signals typically present a first peak (associated to the bounce) followed
by a ring-down. Type II signals show a few (2-3) decreasing peaks, 
with a time lag between the first two  of at least 10\,ms. Type III signals exhibit
no strong peak but fast ($\sim$ 1 kHz) oscillations after the bounce.

As the waveforms in the catalogue are explicitly known, optimal filtering
can be used as a benchmark. The optimal SNR $\rho_0$ 
for a GW signal $h(t)$ (e.g. any of
the 78 signals in the catalogue, located at a distance $d$) is given by :
\begin{equation}
\rho_0^2 = 2 \int {|\tilde{h}(f)|^2 \over S_h(f)} df 
= {f_0 \over \sigma^2} \int |h(t)|^2 dt
\end{equation}
where we use the relation between the one-sided spectral density of the noise
$S_h$, the sampling frequency $f_0$ and the r.m.s of the noise $\sigma$ given by  
Eq. (\ref{sigma}). Since the noise is whitened in the detection
bandwidth, $S_h$ is constant and the Parseval's theorem can be used. The
optimal SNR $\rho_0$ is proportional to $< 1/d >$. We can rather define a (mean)
distance of detection as the distance for which the signal is just detected,
that occurring when $\rho$ reaches the threshold $\tau$. Such mean detection distances have to be found by simulations (because $< d >$ is always larger than  $< 1 / d >^{-1}$). 
With $\tau \simeq 4.89$, we obtain a distance of detection, averaged over the $N_c$ signals
of the catalogue (here $N_c = 78$), $d_0 = {1\over N_c} \sum_{i=1}^{N_c} d_0^{(i)}\simeq 26.5$ kpc,
where $d_0^{(i)}$ is the optimal distance of detection for the i-th
signal of the catalogue. We note that, with the threshold we have chosen,
$d_0$ is of the order of the diameter of our Galaxy; a few signals (those
with large initial rotational energy) have optimal distances of detection
larger than 50 kpc, the distance to the Large Magellanic Cloud. The largest
distance of detection obtained in the sample is about 130 kpc.

\subsection{Definition of the performance}

Let us consider the i-th signal in the Zwerger and M\"uller catalogue. Its
mean optimal distance of detection, defined above, is $d_0^{(i)}$. The mean
distance of detection obtained with another filter is $d^{(i)}$ (averaged
over noise realisations). We then define
the performance of the filter for this signal as $d^{(i)}/d_0^{(i)}$.
This relative definition is convenient, because of the
different ``strengths'' of the signals in the catalogue.
The global performance $\Pi$ of the filter is then defined as the average
of the performances for the $N_c$ signals of the catalogue :
\begin{equation}
\Pi = {1\over N_c}\sum_{i=1}^{N_c} {d^{(i)} \over d_0^{(i)}}.
\end{equation}


Fig. \ref{perfvswindow} shows the performances of the slope and offset
detectors and of ALF, as a function of the window size. The maximal performances are obtained for small window sizes (between 20 and 40 bins, that is 1 ms and 2 ms). The Slope Detector (SD) has a performance greater than $0.6$ for window sizes ($N_{w}$) up to 100 bins (5 ms). The Offset Detector (OD) keeps a performance greater than $0.6$ up to $N = 7.5$ ms while the performance of ALF is always greater than $0.6$ up to $N \simeq 17$ ms. For all the window sizes studied here, all the filters have performances greater than $0.5$. For $X_+$ and $X_-$ (not shown on Fig. \ref{perfvswindow}), the maximal performances are $\Pi_{\mathrm max} \simeq 0.71$ and $\Pi_{\mathrm max} \simeq 0.67$.
The ALF, which combines the informations of the
slope and offset detectors, is the most performant for any window size.


\begin{center}
\begin{tabular}{|c|c|c|c|c|c|c|}
\hline Filter  & SD & OD & ALF\\
\hline Optimal Window Size (ms) & 2.5 & 1.5 & 1.5  \\ 
\hline Performance 	     & 0.65 & 0.70 & 0.78\\\hline
\end{tabular}

~

{Table 1 : Optimal analysis window sizes and performances. The performance is  averaged over the 78 signals 
for all filters, with one single window size for all signals.}

~

\end{center}

\subsection{Window size and detection strategy}

In fact, each of the signals will be optimally detected with a given analysis window size $N_i$ (for signal $i$). The distribution of those window sizes for all ZM signals (see Fig \ref{distrib_win}) shows different ``preferred'' regions. If each signal was detected with its own optimal window size (unrealistic case), the overall performance could rise to $0.91$ for ALF. In fact, to stay as unbiased as possible, it is possible to discretise the window size parameter space, allowing for example 5, 10 or 20 different window sizes used for the same filter. The size of windows and their spacing is not crucial for a given number of window, as the performances of the filters do not depend crucially on these parameters (within typically 1\%).

Of course, the individual threshold for each of the window sizes would have to be higher in order to obtain an {\it overall} false alarm rate (taking into account the redefinition of a false alarm) equal to $10^{-6}$.

\subsection{Performances of the filters}

Table 2 shows the performances obtained with  multi-windowing slope, offset and ALF filters, using $\{$Window Sizes$\}$ = $\{$1.5 ms, 2.5 ms, 5 ms, 10 ms, 15 ms$\}$ (with a clear preference for short duration windows)

We recall also the performances of the Norm Filter (NF) and the Peak Correlator 
(PC) \cite{arnaud,corr}. The Peak correlator is implemented with 26 (truncated) 
Gaussian templates of widths optimally located in the interval [0.1 ms, 10 ms]. 
All filters related to the Slope Detector excepted ALF have performances from 
$0.5$ up to about $0.7$, while ALF reaches a performance greater than $0.8$. 


\begin{center}
\begin{tabular}{|c|c|c|c||c|c|c|c|c|}
\hline Filter  & Optimal & NF & PC & SD & OD & $X_+$ & $X_-$ & ALF\\
\hline Average distance (kpc) & 26.5 & 11.5 & 18.5 & 11.3 & 15.2 & 18.4 & 13.1 & 22.5 \\ 
\hline Performance ($\%$)	   & 1    & 0.46 & 0.73 & 0.49 & 0.59 & 0.66 & 0.54 & 0.81\\\hline
\end{tabular}

~

{Table 2 :Performances of the ALF and related filters, each implemented
with 5 windows in parallel .}
\end{center}

It has been noticed that with 20 window sizes (rather than 5) in parallel, all filters related to ALF have performances around $0.8$. Indeed, to keep an overall false alarm rate of $\kappa_0$ for $n_{window}$ window sizes in parallel, the individual false alarm rate to apply for each window size is roughly given by $\kappa_0 / n_{window}$. This quantity is then tuned by simulations, because each of the filters studied here react in a different way with respect to the event redefinition. The fact that $\Pi_{ALF}^{5 \, windows} \simeq \Pi_{ALF}^{20 \, windows}$ whereas $\Pi_{L}^{5 \, windows} < \Pi_{L}^{20 \, windows}$ (where $L$ denotes all the linear filters, {\it i.e} all the filters except ALF) shows the robustness of ALF with respect to a variation of the detection threshold.

\subsection{Robustness of performances with respect to signal type}

Table 3 shows the mean performances of the filters described above for each of the three types of signal in the ZM Catalogue. Each of the filters are nearly equally performant with type III signals, except $X_+$ and ALF which are much more performant. They all seem to have the same behaviour for type I signals, whereas great differences can be seen in their performances with type II signals (from $0.46$ for SD to $0.72$ for $X_+$).

ALF shows its best performances for type III signals (short durations and high frequencies), whereas types I are preferred by SD and $X_-$, and Type II by OD and $X_+$. Those results give a dispersion (with respect to signal type) of about $5 \%$, for all filters.
\begin{center}
\begin{tabular}{|c|c|c|c|c|c|}
\hline Filter  &  SD & OD & $X_+$ & $X_-$ & ALF\\
\hline Type I signal performance & 0.53 & 0.60 & 0.59 & 0.57 & 0.79\\ 
\hline Type II signal performance & 0.47 & 0.62 & 0.72 & 0.54 & 0.81\\
\hline Type III signal performance & 0.41 & 0.48 & 0.62 & 0.45 & 0.89\\\hline
\end{tabular}

~

{Table 3 : Performances (in percent) of the ALF and related filters. Each filter is implemented
with 5 windows applied in parallel on the different kinds of signals of the ZM Catalogue.}
\end{center}

Concerning the robustness of the filters, one could argue that we have
studied the performance of the filters with only one set of GW signals.
It is worth noting that the linear fit filters have been also tested on 
other ``signals'' than those given by the Zwerger and M\"uller catalogue : generic peaks or damped 
sine\cite{virgonotes}. This kind of signals could be the signature
of typical instrumental artifacts but also of real GW signals such
as black hole oscillations \cite{eche} for example.
The performances in this case are similar (or better) to the benchmark above,
except in the case of high frequency (kHz) and slightly damped
signals (long signals),
 where the performance of ALF, for example, falls down to about 0.3, while
it is close to 1. for very short bursts (Gaussian peaks and strongly
damped sine as well).

\subsection{Efficiency of the filters}

Another way to compare the different filters is to compute their efficiency
as a function of the distance of the source. The efficiency for a given signal located at a given distance is defined as
the number of detections over the total number of simulated noise realisations. Again the efficiency 
is there calculated by averaging over the signals of the ZM catalogue.
Fig. \ref{efficiency} presents the detection efficiency for the
different filters, SD, OD, and ALF as a function of the distance
to the source, expressed in units of the optimal distance of detection for this particular source, and averaged
over all the signals. That means that each source has been located to a distance 
$x \times d^i_{optimal}$, where $d^i_{optimal}$ is the mean detection distance obtained with the Matched Filtering for the $i^{th}$ signal in the Catalogue, and $x \in [0,1]$. The detection efficiency presented here is the mean of the detection efficiencies obtained for each signals in the Catalogue. $X_-$ (not shown on Fig \ref{efficiency}) behaves like OD at small distances and like SD at larger distances (this is the contrary for $X_+$).

We can also derive the distances $d_{e}$ for which each signal reaches a  efficiency 
of $e$\%(see Table 4). We note that in spite of performances rather different for all 
the filters presented here, their detection efficiencies behave quite similarly.
Eventually, all filters have the same {\it effective performance} $\Pi_{\mbox{eff}}$ 
defined by $d_{50}/d_{\mathrm optimal}$, which is around $0.75$.
We note also that the efficiency of the Wiener filter is around 50\% for the optimal 
distance of detection $d/d_{\mathrm optimal} = 1$.



\begin{center}
\begin{tabular}{|c|c|c|c|c|c|c|}
\hline Filter  & Optimal & SD & OD & $X_+$ & $X_-$ & ALF\\
\hline $d_{95}/d_{\mathrm optimal}$ 	 & 0.65    & 0.47  & 0.5 & 0.47 & 0.49 & 0.52\\
\hline $d_{50}/d_{\mathrm optimal}$ 	 & 0.96       & 0.72 & 0.73 & 0.72 & 0.73 & 0.75\\\hline
\end{tabular}

~

{Table 4 : Average source distances for which the ALF and related filters 
reach a 95\% (50\%) efficiency}
\end{center}





Figure 7 shows the false dismissal rate (ratio of missed events or {\it inefficiency}) as a function of the false alarm rate, for signals located at 10 and 25 kpc, for a single windowed ALF filter. At $10 \mbox{kpc}$, with realistic false alarm rates (of the order of $10^{-6} - 10^{-7}$), the dismissal rate is about $40 - 50 \%$. Small dismissal rates can be reached with extremely high false alarm rates (greater than $10^{-3}$). Even if the performances of such a filter seems to be very high, the detection of sources (from such a catalogue) out of our Galaxy will be clearly difficult. The last curve (labelled $d_{95}$) shows the evolution of the false dismissal rate where each signal of the Catalogue has been located to a distance at which, for final ALF (5 windows), the detection efficiency is $95 \%$. This means that in this case, each source is located at the same fraction of its own optimal detection distance (hence a {\it different} distance for each source).

\section{Sensitivity of the Filters to Non-White Noise}

It is likely that the data provided by the interferometric detectors will be non-Gaussian and non-white. On-line filters will process pre-whitened data, and this whitening will be certainly non perfect. This is a crucial point to know how online triggers will behave in such a case. In order to study this effect, we thus have added to the white Gaussian noise low frequency components of the type  $A_{bf} sin(2 \pi f t)$. We then compute, as a function of the frequency $f$, the amplitude $A_{10 \%}$ which corresponds to an increase of $10 \%$ of the number of false alarm, for the final multi-windowing ALF filter.


For $f = 0.6 Hz$ (pendulum mode in Virgo), we find that the maximum authorised amplitude is of the order of $2 \times 10^{-2} \times \sigma_{noise}$. For other frequencies, $A_{10 \%}$ is in the range $2 \times 10^{-2}$ - $5 \times 10^{-2}$ ($\times \sigma_{\mbox{noise}}$) up to 1 kHz. Fig. 9 shows the evolution of the false alarms excess as a function of the amplitude of the $0.6 Hz$  component in imperfectly whitened samples of data. A low frequency amplitude about 5 times larger than $A_{10 \%}$ for $f = 0.6 Hz$ gives 10 times more false alarms. This proves (if needed) that the whitening process will be a crucial part of the analysis procedure.


%
%

\section{Conclusions}

We have designed and tested some filters based on linear fits and aimed at the
 detection of short bursts of gravitational waves, such as the ones
emitted by massive star collapses. In particular, we have built a non linear
filter, ALF, from the slope and offset detectors resulting from the fit procedure. These filters match the simplicity and
speed requirements needed for on line triggers to be easily implemented in 
interferometric detectors of gravitational waves.
The performances of the filters are better than those of the
filters studied in \cite{arnaud} with the same procedure.
In particular, the ALF filter, if implemented with 5 windows, reaches
a performance of about $80 \%$, relative to the matched filter, and the detection efficiency for such signals is about $50 \%$ at a distance $d \sim 0.75 \times d_{\mbox{optimal}}$.
This is, however, just enough to detect supernova
signals from anywhere in the Galaxy. Indeed, the mean distance of detection
averaged over the supernova signals contained in the Zwerger and M\"uller catalogue
is about 22.5 kpc, of the order of the diameter of the Galaxy. This figure
has been obtained by assuming an optimal incidence of the (``+'' polarised)
signals along
the arms of the interferometric detector, and should be in fact corrected
for the incidence effect. Averaging over all the possible source locations
in the sky reduces the signals by at least $1/\sqrt{5}$ (in the isotropic case).
 This finally results in a mean distance of detection of about 10 kpc.
Clearly, with the low expected rate of supernovae in our galaxy
(about 3 per century) massive star collapses are not likely to 
be detected with the
first generation of interferometric detectors, unless the asymmetry of
the collapse is much larger than presently expected in current models.
It is not impossible, however, that the first generation detectors
could be sensitive to binary mergers as far as the Virgo cluster
(especially black hole mergers). Indeed, rough estimates of the amplitudes of 
the GW signals are two or three orders of magnitude larger than the
ones predicted for the supernova signals; this gives crudely distances
of detection that might be as large as 10 Mpc with the present detectors,
LIGO I and Virgo. Regarding the detection of GW emitted by a binary
system, combining the traditional method of matched filtering
for the inspiral phase and a ``pulse'' detection technique, such 
as those developed in this paper and in \cite{arnaud,moriond,pra},
could help to increase the final
signal to noise ratio and then the confidence in the detection
of an interesting event. This can be valuable especially for binary black
holes, for which only a few cycles can span the detector bandwidth,
so that the contribution of the merging phase to the total signal to noise
ratio can be important (see \cite{bd} for an idea of the
respective strengths of the inspiral and merger waveforms).

Finally, a filter such as ALF seems to fulfil all requirements
 to be implemented as part as on line
trigger dedicated to detect bursts of unknown waveform
in interferometric detectors of gravitational waves.

\baselineskip = 0.5\baselineskip  

 
\newpage


\begin{figure}
\centerline{\epsfig{file=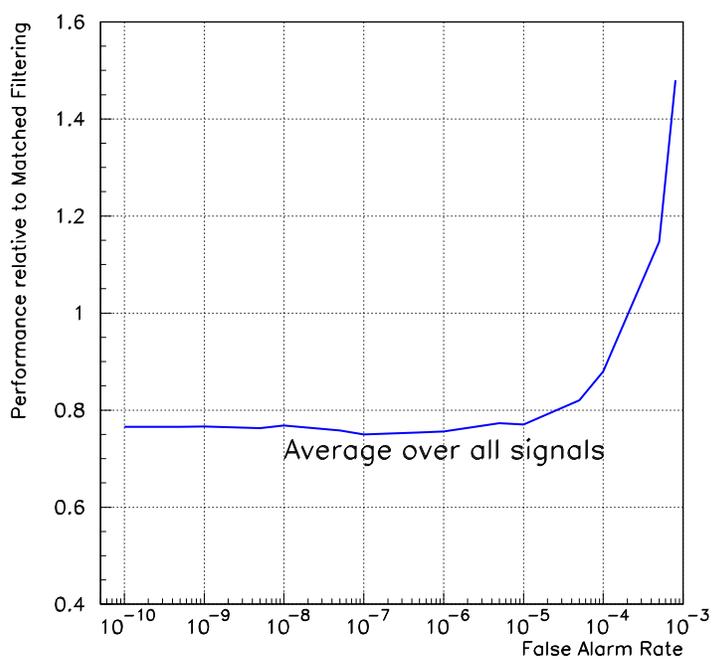,height=10cm,width = 10cm}} 
\caption{Relative Performance of ALF (single-windowed) as a function of the False Alarm Rate, averaged over all the signals of the catalogue.}
\label{perfvskfa}
\end{figure}

\begin{figure}
\vskip -1cm
\centerline{\epsfig{file=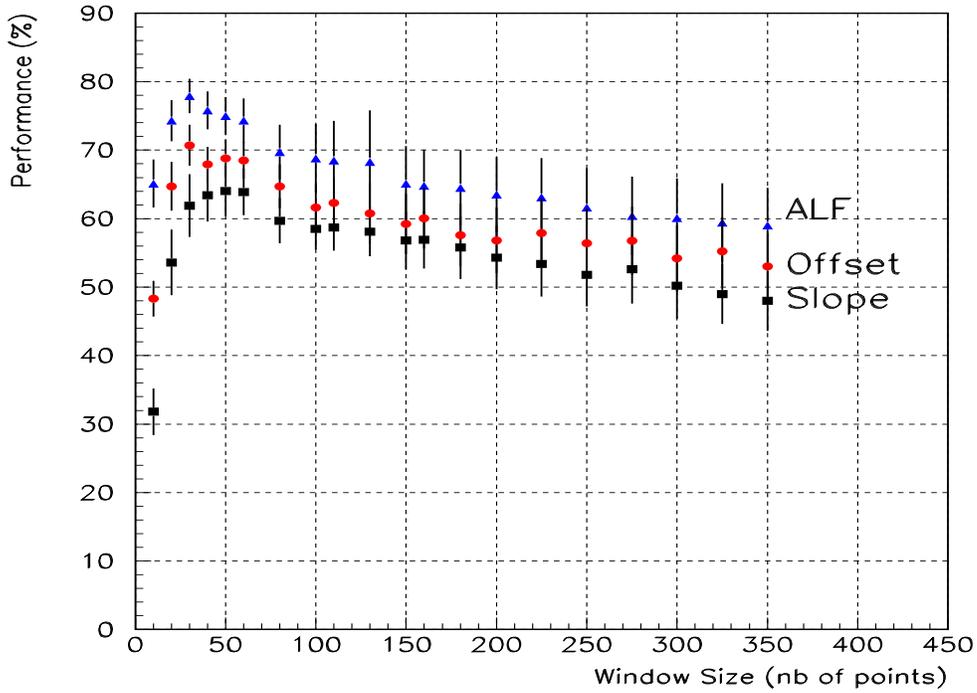,height=14cm,angle=270}} 
\caption{Performance of Single-Windowing Filters Slope, Offset and ALF as a function of the window size. The error bars take into account the finite statistic of the waveforms taken from the Catalogue.}
\vskip -1.4cm
\label{perfvswindow}
\end{figure}

\begin{figure}
\centerline{\epsfig{file=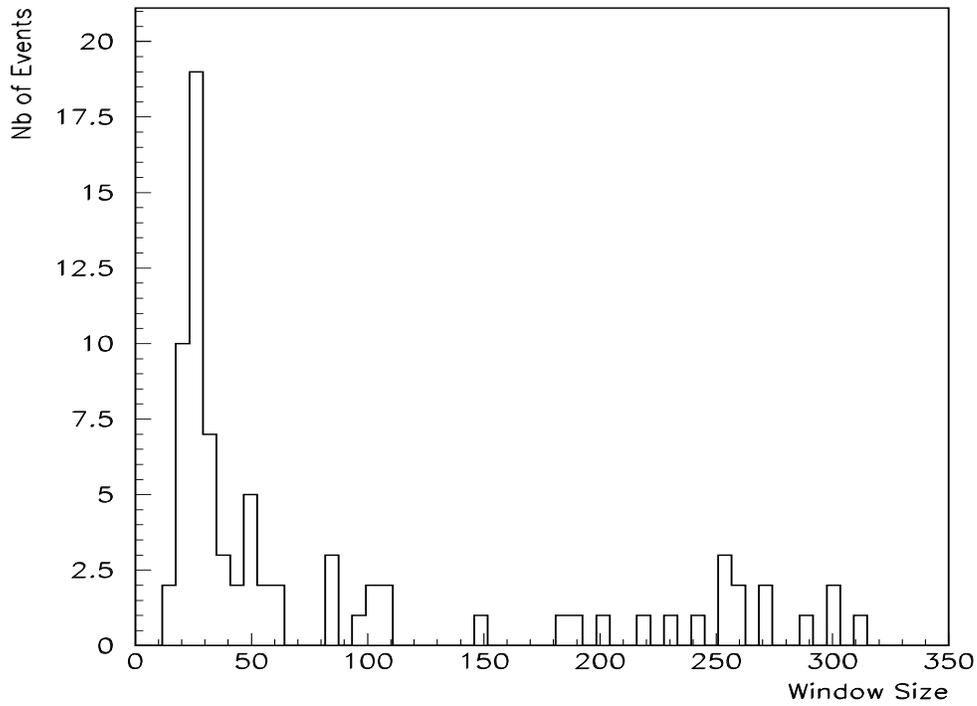,height=14cm,angle=270}} 
\caption{Distribution of Window Sizes (in number of samplings at 20 kHz) that give optimal performances for ALF. Short duration windows (up to about 100 bins, i.e 5 ms) are clearly preferred.}
\vskip -1cm
\label{distrib_win}
\end{figure}

\begin{figure}
\vskip -1cm
\centerline{\epsfig{file=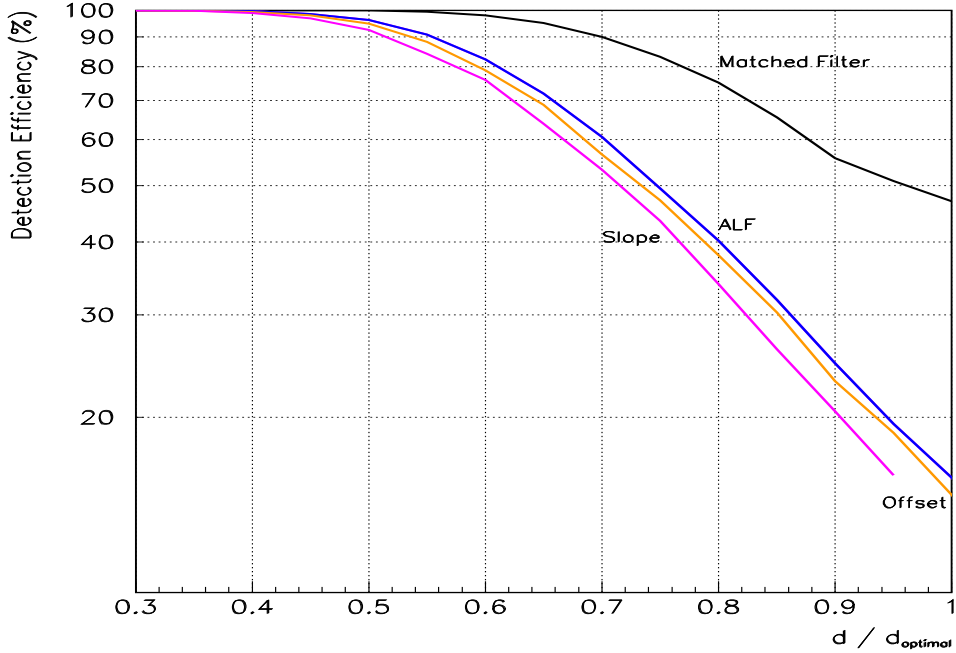,height=10cm,width=14cm}} 
\caption{Detection efficiency of the filters as a function of the distances
of the source.The distances are normalised to the optimal distances of 
detection for each signal and the efficiency is averaged over all the signals
of the Zwerger and M\"uller catalogue.}
\vskip -1.05cm
\label{efficiency}
\end{figure}
\begin{figure}
\centerline{\epsfig{file=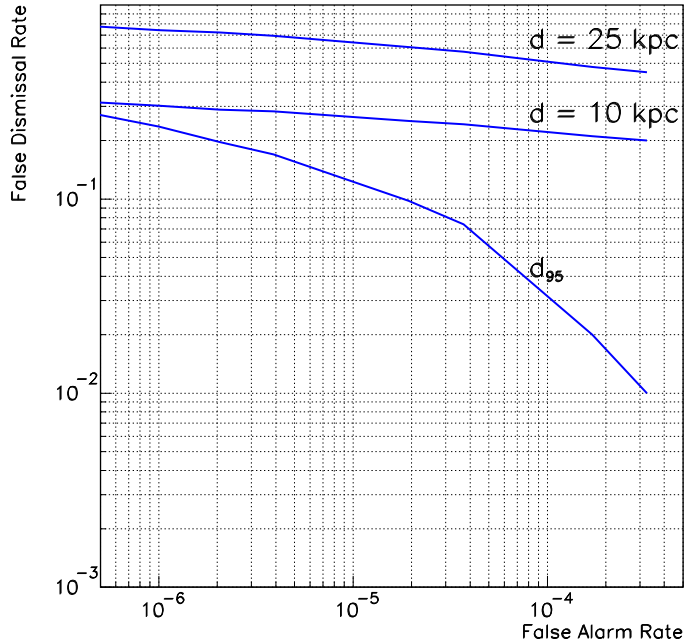,height=10cm}}
\vskip 0cm 
\caption{False Dismissal Rate as a function of False Alarm Rate for all signals located at 10 kpc and 25 kpc. The curve labelled $d_{95}$ concerns signals located at a distance such that for ALF (with 5 different window sizes), $\epsilon_{ALF} \simeq 95 \%$ (for $\kappa_0 \sim 10^{-6}$), that is, a {\it different} distance for each signal. Dashed lines represent 25 $\%$, 50 $\%$ and 75 $\%$ false dismissal rate.}
\label{farate}
\end{figure}
\begin{figure}
\vskip -1cm
\centerline{\psfig{file=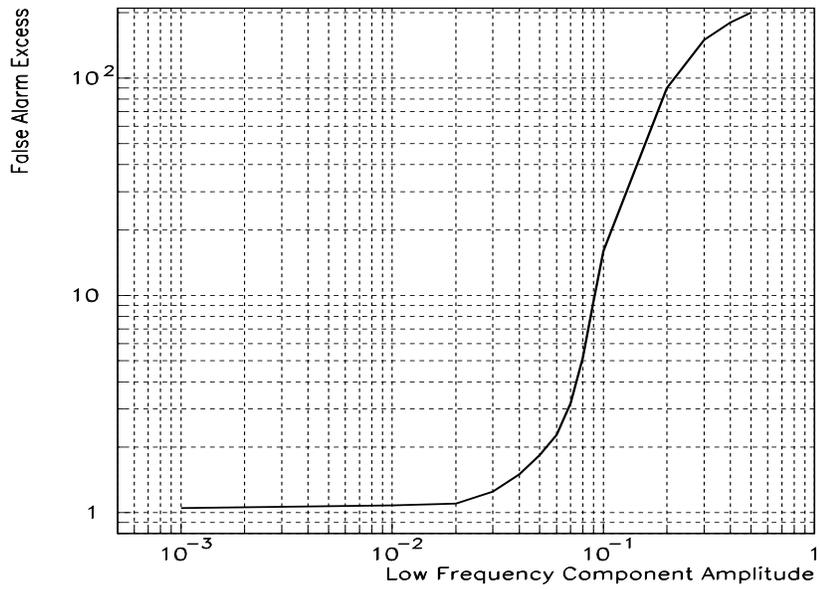,height=12cm,angle=270}}
\caption{False Alarms Excess as a function of the amplitude of a low frequency component (here 0.6 Hz) added to white noise. This excess is the quantity (Effective Number of False alarms)/(Allowed Number of False Alarms). The amplitud8e is measured relatively to the standard deviation of the noise. In this figure, the amplitude $A_{10\%}$ represents a false alarm excess of $1.1$.}
\label{ampvsfa}
\end{figure}



\begin{references}
\bibitem{corr} The $\sqrt{2}$ factor forgotten in \cite{arnaud} has been here
corrected for.
\end{references}
\end{document}